\documentclass[page-classic]{epl2} 

\usepackage{amsmath}
\usepackage{amssymb}
\usepackage{graphics}
\usepackage{epsfig}
\usepackage{color}

\newcommand{\be}{\begin{equation}}
\newcommand{\ee}{\end{equation}}

\newcommand{\cmod}[1] {|{#1}|^2}
\newcommand{\vr}[1] {{\vec r_{#1}}}
\newcommand{\vR}[1] {{\vec R_{#1}}}

\title{Spontaneous Polaron Transport in Biopolymers}

\author{B. Chakrabarti\inst{1}
  \thanks{E-mail: \email{buddhapriya.chakrabarti@durham.ac.uk}}, 
B. M. A. G. Piette\inst{1}
  \thanks{E-mail: \email{b.m.a.g.piette@durham.ac.uk}} \and 
W. J. Zakrzewski\inst{1}
  \thanks{E-mail: \email{w.j.zakrzewski@durham.ac.uk}}}
\shortauthor{B. Chakrabarti \etal}

\institute{
  \inst{1} Department of Mathematical Sciences, Durham University, 
Durham, DH1 3LE, United Kingdom.\\
}

\abstract{Polarons, introduced by Davydov to explain energy transport
in $\alpha$-helices, correspond to electrons localised on a few lattice sites
because of their interaction with phonons. While the static polaron
field configurations have been extensively studied, their displacement
is more difficult to explain. In this paper we show that, when the
next to nearest neighbour interactions are included, for physical
values of the parameters, polarons can spontaneously move, at $T=0$, on bent 
chains that exhibit a positive gradient in their curvature. At room temperature
polarons perform a random walk but a curvature gradient can induce a non-zero 
average speed similar to the one observed at zero temperature. 
We also show that at zero temperature a polaron bounces on sharply kinked 
junctions. We interpret these results in light of the energy transport by
transmembrane proteins.}

\begin{document}

\maketitle

\section{Introduction} Proteins, essential components of all
biological cells are central to their proper functioning. As ``form
determines  function'', a study of protein structure and dynamics is
of utmost importance in elucidating their role in cellular
behaviour. One of the key problems in biology is to understand energy
transport from one part of the cell to another  and to study the role
of protein conformations and conformational transitions in this
process.

The mechanism of charge and energy transport in proteins and other
bio-macromolecules at the atomic scale was proposed  Davydov and
co-workers\cite{Davydov}. In this approach the transport properties
are considered in terms of the emergence  of a `polaron' whose
properties and dynamics are used to describe the resultant
transport. The polaron describes a localised excitation which carries
energy corresponding to some vibrational modes of a group of
molecules and a distortion of the chain containing these
molecules. The system acts as a particle and its dynamical  properties
can be studied in terms of solutions of particular differential
equations that describe its behaviour.

The Davydov theory hinges on the assumption that an extra electron or
energy quanta released in the hydrolysis of  ATP (adenosine
triphosphate) can be stored by the protein molecule in its vibrational
mode. The non-linear coupling  between the vibrational mode and other
excitations on the chain leads to the formation of a soliton (polaron
in the case of the interaction between the phonon modes and the
electron in a polarizable medium). The soliton (polaron in this case)
can then propagate along the polypeptide backbone leading to
energy/charge transport. The soliton mediated transport  mechanism has
been applied to helical proteins\cite{Davydov:85}. The theory of
non-linear energy transport in bio-molecules  is reviewed in
\cite{Scott:92,Cruzeiro:09}. Most of the studies are based on simple
models of one-dimensional reductions  of the three-dimensional
structure of the $\alpha$-helix proteins \textit{i.e.}, only a single
strand of hydrogen-bonded  peptide units is considered. Moreover, the
original set of discrete lattice equations is often treated in a
continuum  approximation. Recent studies based on two
and three-dimensional models describing the  solitonic and/or
polaronic  transport of energy are described in detail  in
\cite{Olsen:88,LaMagna:95,Zolotaryuk:96,Christiansen:97}.

Studies of polaron transport on $\alpha$ helical polymers have
recently  been reported by Henning~\cite{Henning:02}. The Henning
model is an extension of semiclassical treatments of polaron transport
within the Holstein model which  incorporates three strands that
comprise the $\alpha$ helix. Numerical schemes and variational
approaches have been utilized in the  treatments of the Holstein model
with a hard non-linear on-site potential term in one dimension to
obtain polaronic ground  states, their phase diagram in terms of the
non-linearity parameter as well as the one characterising the
electron-phonon  coupling\cite{Kalosakas:98, Voulgarakis:01,Eremko04}. It has
also  been shown that the presence of the non-linear on-site potential 
polaron formation is possible in higher dimensions, a fact borne
out by the  Henning  model. Normal mode analysis of this model
revealed the existence of a low-frequency pinning mode, and hence
enabled the construction of one and two-dimensional moving  polaron
solutions by a Floquet analysis\cite{Kalosakas:98,
Voulgarakis:01}. The effects of the lattice non-linearity were
reported to have led to a  dramatic reduction of the polaron effective
mass.  

However, in most of these studies the polaron is ``kicked'' from its
rest state via a perturbation\cite{Brizhik10}. In this paper we  look at the
spontaneous polaron transport via charge-conformational coupling. In
particular, we study polaron transport  on a flexible chain with an
imposed initial bend at both zero ($T=0$) and non-zero ($T\ne0$)
temperatures. We now  summarise our main results. 

At $T=0$ a polaron can undergo spontaneous motion via the coupling of
charge with the conformational degrees of freedom. Thus  an imposed
bend on the chain causes the polaron to accelerate. However,  we have
found that when such an accelerating polaron  encounters a kink, a
slope discontinuity along the chain backbone, it gets reflected instead
of continuing along its  original direction of motion guided by
inertia. The reason for this due  to the long-range nature of the
exchange coupling term. At finite temperature $T\ne0$ thermal
fluctuations wash away directed transport and the polaron  undergoes
large amplitude fluctuations about its mean position. However, since
the chain ends act as reflecting walls, a  polaron formed closer to
one end of a straight chain would often get reflected, resulting in an
overall small but  non-negligible drift.

The paper is organised as follows: In the next section we present the
Hamiltonian describing polaron transport on a flexible chain  and
change variables to dimensionless quantities.  The dimensionless
parameter values corresponding to  physically relevant systems
\textit{e.g.} proteins are computed in Sec.3. We describe the  results
of our numerical studies in Sec.4  and we set our work in perspective
in the final section. 

\section{Polaron Hamiltonian on a Flexible Chain}
\label{PolaronModel} We consider a modified version of the polaron
model proposed by Mingaleev \textit{et al.}\cite{Ming02}. Our model
involves a semi-classical treatment of the interaction between a
phonon field $\vR{n}$ and an electron field $\phi_n$ on a flexible
linear chain whose  nodes are  labelled by an index $n$.

The Hamiltonian of the model is given by
\begin{eqnarray}
H &=& \sum_{n} \left[\frac{\hat{M}}{2} \left(\frac{d \vR{n}}{d\tau}\right)^2
   + \hat{U}_n(\vR{})
   +  W\left(2\cmod{\phi_n} - \sum_{m\ne n}J_{nm}\phi_n^*\phi_m \right)
            - \frac{1}{2} \Delta |\phi_n|^4
\right],
\end{eqnarray}
where $M$ is the mass of the chain node, $W$ is the linear excitation transfer 
energy and $\Delta$ the non-linear self-trapping interaction. The excitation transfer 
coefficients $J_{n,m}$ are of the form:
\begin{equation}\label{alpha-Eq}
J_{n,m} = J(|\vr{n}-\vr{m}|) = (e^\alpha -1)\,
           e^{-\alpha|\vr{n}-\vr{m}|/a},
\end{equation}
where $\alpha^{-1}$ sets the relative length scale over which the interaction 
decreases, in units of $a$, and where $a$ is the rest distance between two 
adjacent sites. The $J_{n,m}$ describes the long range interaction between the 
electron field at different lattice sites $n$ and $m$;  its value decreases 
exponentially with the distance between them.

Note that  the normalisation of the electron field is preserved in our model, 
\textit{i.e.}
\begin{equation}
\sum_n |\phi_n|^2 = 1.
\end{equation}

The phonon potential $U_n$ consists of three terms:
\begin{eqnarray}
\hat{U}_n(\vR{}) = \frac{\hat{\sigma}}{2} (|\vR{n}-\vR{n-1}|-\hat{a})^2 +
\frac{\hat{k}}{2} \frac{(\theta_n-\varphi_n)^2}
                       {\left[1-((\theta_n-\varphi_n)/\theta_{max})^2\right]} \nonumber \\
          + \frac{\hat{\delta}}{2} \sum_{m\ne n} (\hat{d}-|\vR{n}-\vR{m}|)^2
                      \Theta(\hat{d} -|\vR{n}-\vR{m}|).
\label{defUphys}
\end{eqnarray}

The first term in Eq.(\ref{defUphys}) models the elastic energy
describing the  stretching of the adjacent nodes of the chain, where
$\hat{a}$ is the equilibrium separation between them. The second term describes the 
bending energy of the chain akin to
semi-flexible polymers. Here $\varphi_n$ is the rest angle between
adjacent lattice links and $\theta_{max}$ is the  largest angle
allowed. In the Mingaleev model\cite{Ming02} $\varphi_n=0$ and the
equilibrium configuration of the chain  is a straight one. Non-zero
values of $\varphi_n$ lead to a bent chain. The last term,
(proportional to  $\hat{\delta}$), models hard-core repulsion between
the atoms of the chain. However, we note that this term does not
contribute significantly in determining the equilibrium chain
conformation.

In this paper symbols denoted by an overhead carat sign \textit{e.g.}
$\hat{M}$, $\hat{\sigma}$ \textit{etc.} correspond  to physical
variables carrying units and dimensions while those without it
correspond to non-dimensional variables and  parameters. 
We rescale time $\tau$ by defining a timescale $\tau_0=\frac{\hbar \Delta}{W^2}$s 
and rescale distances by a length scale $a$. The non-linear coupling 
parameter $g$ appears as a dimensionless ratio of two energy scales.  
\begin{eqnarray}
\tau &=& t \tau_0
\qquad\qquad
g = \frac{\Delta}{W}
\qquad\qquad
r = \frac{R}{a}.
\end{eqnarray}

In terms of these variables the Hamiltonian takes the form
\begin{eqnarray}
H &=& \frac{W^2 }{\Delta} \sum_{n}
 \left[ \frac{M}{2}\left(\frac{d\vr{n}}{dt}\right)^2
   +  U_n(\vr{})
   +  g\left(2\cmod{\phi_n} - \sum_{m\ne n}J_{nm}\phi_n^*\phi_m \right)
            - \frac{g^2}{2} |\phi_n|^4
\right],
\label{eqH}
\end{eqnarray}
where
\begin{equation}
U_n(\vr{}) = \frac{\sigma}{2} (|\vr{n}-\vr{n-1}|-a)^2 +
\frac{k}{2} \frac{(\theta_n-\varphi_n)^2}
                 {\left[1-((\theta_n-\varphi_n)/\theta_{max})^2\right]}
           + \frac{\delta}{2} \sum_{m\ne n} (d-|\vr{n}-\vr{m}|)^2
                      \Theta(d -|\vr{n}-\vr{m}|)
\end{equation}
with
\begin{eqnarray}
M &=& \hat{M} \frac{a^2 W^2}{\hbar^2 \Delta}
\qquad\qquad
\sigma = \hat{\sigma}\frac{a^2\Delta}{W^2}
\qquad\qquad
\delta = \hat{\delta}\frac{a^2\Delta}{W^2}\nonumber\\
k &=& \hat{k} \frac{\Delta}{W^2}
\qquad\qquad
a = \frac{\hat{a}}{a} = 1
\qquad\qquad
d = \frac{\hat{d}}{a}.
\end{eqnarray}

Writing $\vr{n}=(x_{1,n},x_{2,n},x_{3,n},)$ we can derive the equation 
of motion for $x_{i,n}$ from the Hamiltonian 
Eq.(\ref{eqH}). Thermal fluctuations are incorporated by adding a delta 
correlated white noise $F(t)$ which satisfies 
\begin{equation}
<F(0) F(s)> = 2 \Gamma k_B T \delta(s).
\end{equation}
The resulting Langevin equations describing 
the coupled dynamics of the chain and of the 
polaron is given by
\begin{eqnarray}\label{Polaron-Chain-Dynamics-Eq}
M\frac{d^2x_{i,n}}{dt^2} + \Gamma \frac{d x_{i,n}}{dt} + F(t)
  +\sum_m \frac{dU_m}{dx_{i,n}}
  -g\sum_k \sum_{m < k}\frac{d J_{km}}{dx_{i,n}}
               (\phi_k^*\phi_m+\phi_m^*\phi_k) &=& 0
\nonumber\\
i\frac{d\phi_n}{dt}-2 \phi_n + \sum_{m\ne n} J_{nm}\phi_m
   + g \cmod{\phi_n}\phi_n
   &=& 0,
\end{eqnarray}

Note that we also have
\begin{equation}
\overline{k_{B} T} = k_B T \frac{W^2}{\Delta}= k_B T W g
\end{equation}
and, as the equation for $x_i$ is expressed in units of $W^2/(\Delta a)$,
we have $\Gamma = \hat{\Gamma} a^2/\hbar$.


\section{Physical Parameter Values}\label{ParameterValues}

In order to study the feasibility of the mechanism of the energy transport via 
polarons for biologically relevant systems 
we have used the parameter values corresponding to those of $\alpha$-helices. 



We note  that for Amid-I vibrations in $\alpha$-helices \cite{Scott82} we have:
$W\approx 2\times 10^{-22}J\approx 1.2meV$, $\hat{\sigma}=19.5N/m$ and
$\hat{M}=2\times 10^{-25}kg$. $a \approx0.45nm$.
$\epsilon=0.02eV$.
$\Delta=8\frac{\epsilon^2}{\hat{\sigma}}=4.74\times 10^{-22}J=0.003eV$. 
Moreover, $\hat{k}$ can be evaluated from the 
persistence length of $\alpha$-helices $\lambda\approx 65 nm$ 
\cite{Phillips1996}
\begin{equation}
\hat{k} = \lambda k_B T /a \approx 6\times 10^{-19}J.
\end{equation}
Though we do not have experimental values of $\hat{\delta}$, it is clear that 
it must be larger than $\hat{\sigma}$. For 
the friction coefficient we assume that $\hat{\Gamma} \approx 6\pi\mu a $,
 where $\mu= 0.001 Pa\,s$ is the water 
viscosity. Incidentally, for the cytoplasm $\mu$ is up to $4$ times larger 
than this value.

At ``room'' temperature $T=300K$, the non-dimensional parameter values are
\begin{eqnarray}
g = 2.5 \qquad\qquad&&\sigma = 51351.\qquad\qquad k = 7776
        \qquad\qquad\Gamma = 8143\\
M=2.79\times 10^{5}\qquad\qquad &&k_BT=54\qquad\qquad \nu=0.01
\qquad\qquad \tau_0 = 10^{-12}s.
\nonumber
\end{eqnarray}
Following Mingaleev \textit{et al.} \cite{Ming02} we choose $\alpha=2$.

Next we have  performed the simulation of the time evolution described
by the equations Eq.(\ref{Polaron-Chain-Dynamics-Eq})  - to explore the
coupled dynamics of polaron transport  and the conformational
transitions of the chain associated with such motion. For the values
of $g$ and $k$ mentioned  above, we have found that the physical
$\alpha$-helices are too rigid for the polaron to bend the chain and
move along it as suggested for DNA in \cite{Ming02}  (see Fig. 2 
in \cite{Ming02}  where our $g$ corresponds to $N$). 

As a matter of fact, physical polarons always have a relatively 
small energy, too small to bend a polymer spontaneously, even for DNA. 
Nevertheless, the Mingaleev et al. model can be used to study the properties 
of polarons on chains that are, like most proteins, naturally bent. 
One can expect the electron-phonon interaction to lead to the spontaneous 
displacement of polarons on bending gradients for the following reason: 
the main  effect of the electron-phonon interaction is
to favour configurations where the distance between nodes is reduced
(\textit{i.e.} the interatomic separation is smaller than the
equilibrium lattice spacing in the vicinity of the polaron).  As a
kink in the linear chain reduces the distance between nodes that are
not adjacent to each other, one expects that  the polaron would be
attracted by the kink. Thus chain conformations where the angle
between consecutive tangent vectors  increases monotonically serve as
an attractive potential as seen by the polaron. It is thus expected
that the polaron will  accelerate towards a region of high curvature.

In the next section we present the results of  investigations of
whether these expectations are correct for the physically relevant
proteins  ({\it i.e} for chains with physically relevant values of their
parameters).

\section{Polaron on a Bent Chain}\label{Results}

We consider a chain of $N=60$ points with a bent mid-section \textit{i.e.} 
the region between $n=25$ and $n=45$. Such 
a configuration is achieved by setting, initially, $\varphi_n$ as follows:
\begin{equation}
\label{a}
 \begin{array}{ll}
  n< 25 & \varphi_n=0 \\
  25 \le n \le 45 & \varphi_n= (n-25) d\varphi \\
  n > 45 &\varphi_n=0, \\
  \end{array}
\end{equation}
where $d\varphi$ is the incremental increase in the angle between adjacent 
links.  

In order to numerically simulate the dynamics of the polaron on such
a chain, we initially generated and saved the polaron on an undeformed
(\textit{i.e.} straight) lattice ($\varphi_n = 0$ for all $n$). Such a polaron 
was obtained by a relaxation method: a friction term was added to the phonon 
field and the eigenvalue problem for the electron field was solved simultaneously by 
relaxation. Starting with an electron spread over a few lattice 
sites and located on a node which we have chosen to be $n = 30$, the 
system was relaxed until both the electron and the phonon fields reached a 
stationary configuration. This produced a polaron 
configuration centred at $n = 30$. The lattice was then deformed by changing 
the values of $\varphi_n$ to those of (\ref{a}) and a second  
relaxation was performed to let the phonon field reach its equilibrium configuration. The 
electron field of the relaxed polaron was then restored, the time set to 
$t=0$, and the equations for the phonon and electron fields integrated 
numerically.

The zero temperature ($T=0$) dynamics was obtained by numerically
integrating the electron and phonon fields in
Eq.(\ref{Polaron-Chain-Dynamics-Eq}) employing a fourth order
Runge-Kutta scheme, noting the time it
took for the electron to start moving.  

\begin{figure}[ht]
\unitlength1cm \hfil
\begin{picture}(14,7)
 \epsfxsize=7cm
  \epsffile{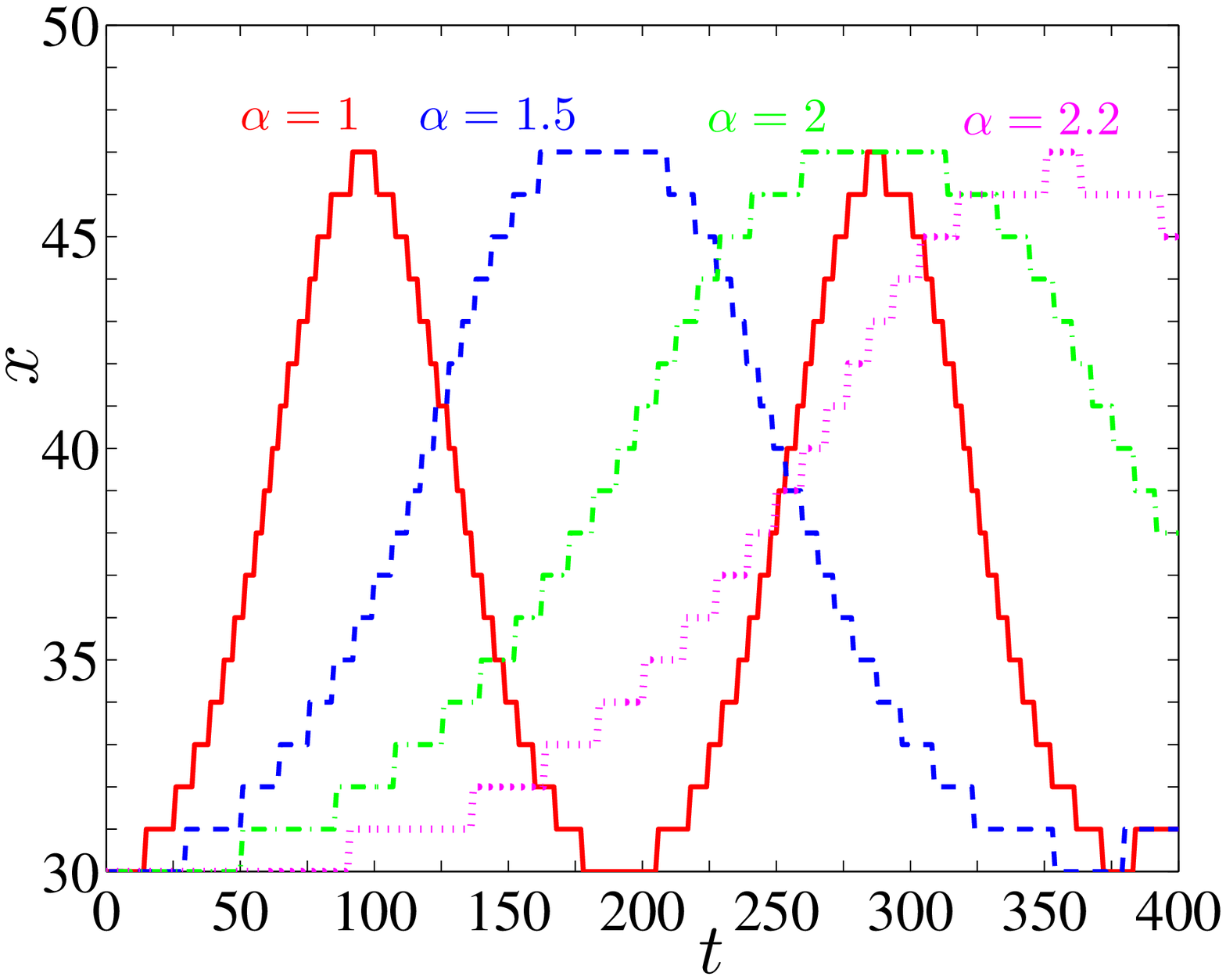}
 \epsfxsize=8cm
  \epsffile{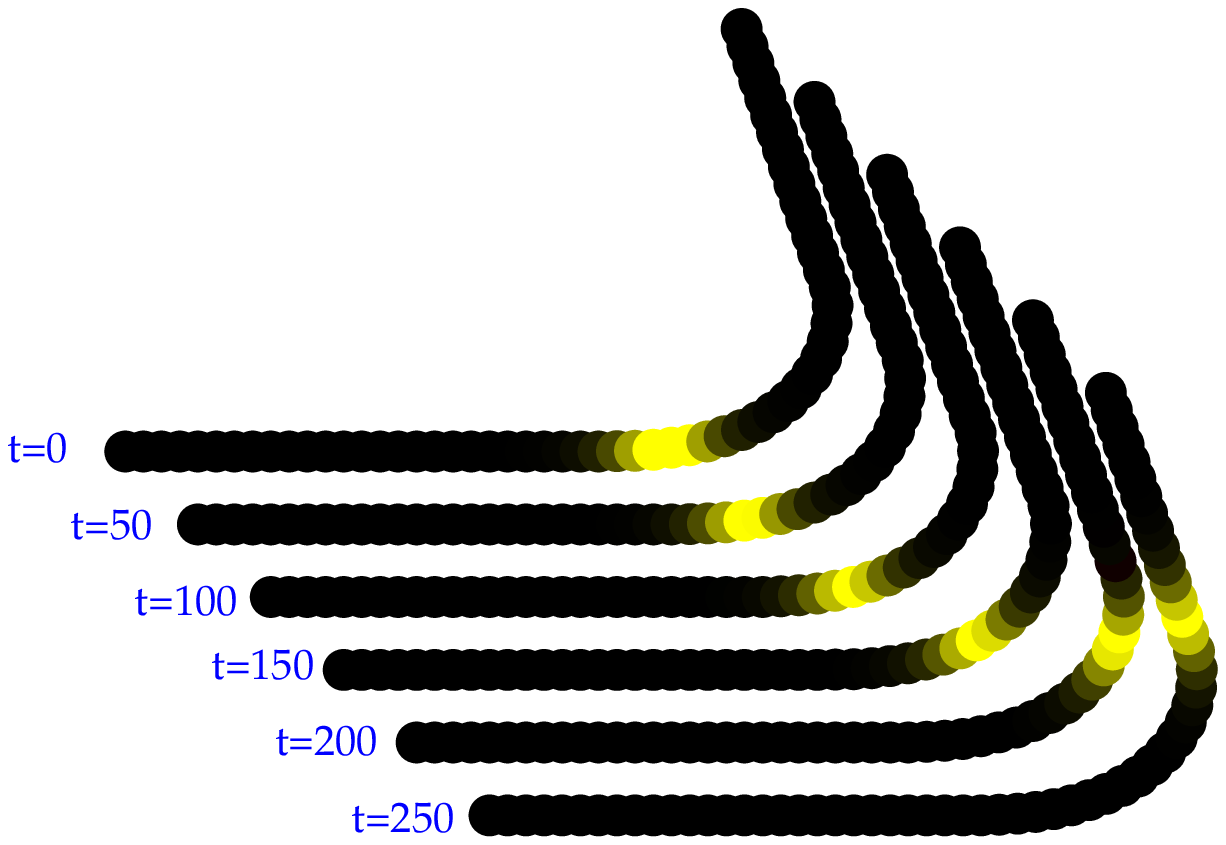}
\end{picture}
\caption{Polaron displacement on a bent chain with $d\varphi=0.01$ at $T=0$ 
for a $N=60$ chain with the initial polaron position at $x_0=30$. The time 
step of integration is chosen to be $dt=50$. Panel (a) shows the trajectory 
of the polaron on the bent chain for different values of $\alpha$ ($\alpha=1$ 
(red solid line), $\alpha=1.5$ (blue dashed line), $\alpha=2$ (green dash-dotted 
line), and $\alpha=2.2$ (magenta dotted line). Panel (b) shows the chain configuration 
and the electron density. Each circle corresponds to a node and a lighter colour 
corresponds to a higher electron density.}
\label{Fig1}
\end{figure}

Fig. \ref{Fig1} summarises our results for $T=0$. Panel (a) shows
typical trajectories of a polaron for different values  of $\alpha$,
the parameter that models the range of the long-ranged interaction
$J_{mn}$ (Eq.(\ref{alpha-Eq})) Our  simulations show that the main
effect of including the next to the nearest neighbour interaction is
to make the polaron  ``to get attracted'' to the bend. This is caused
by the fact that the lattice points near this bend are nearer to each
other  and this lowers the energy of the polaron. Hence, on a bent
chain, with a slowly increasing bending angle, the polaron is
attracted increasingly to the bend as it moves along the chain. This
corresponds to the spontaneous displacement of the  polaron. It is
curious to note that upon reaching the edge of the bent chain beyond
which the chain is straight the  polaron is reflected. This can be explained 
by the fact that the energy of the polaron is the lowest where the bending 
angle is large and the highest where the chain is straight. The transition 
from a bent to a straight configuration thus 
corresponds to a potential wall on which the polaron bounces.
One notices also that as the parameter $\alpha$ is increased the exchange 
energy between non adjacent nodes decreases more rapidly with the distance 
separating them.  This results in a smaller acceleration of the polaron as seen 
in Fig. 1. The speed of the polaron was evaluated as 
$V = [n(t_0+\Delta t)-n(t_0)]/\Delta t$ where $t_0$ is the time at which the 
polaron started to move and $\Delta t=50$. When the polaron travelled 
a distance larger than 20, we took $V = 20/\Delta t$ where $\Delta t$ was
the time taken to cover this distance.

Fig. \ref{Fig1}(b) shows the chain configuration with the electron
density superimposed on it. Each filled circle (black)  corresponds to
a lattice point, while the electron density, denoted in grey/yellow, is 
overlayed on them. The intensity of the lighter circles  provides a measure of
the electron density at a particular site, \textit{i.e.} a lighter
colour corresponds to a higher  electron density. The maximum in the
electron density corresponds to the position of the polaron used to
generate  Fig. \ref{Fig1}(a). 


\begin{figure}[tbp]
\centerline{\includegraphics[scale=0.4]{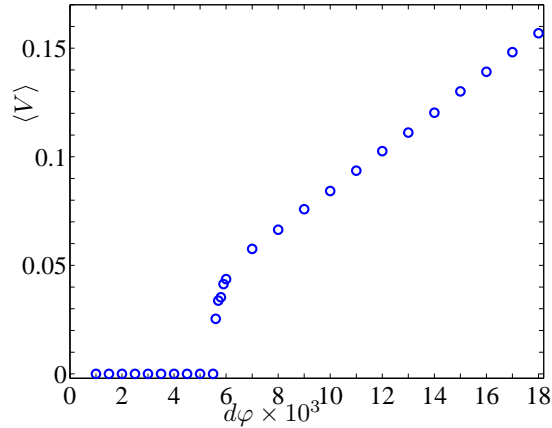}} \vspace*{-0.3cm}
\caption{Figure showing the variation of the velocity $\langle V \rangle$ as a 
function of the bend angle $d\varphi$ for a bent chain of $N=60$ nodes. 
The initial position of the polaron is at $x_0 = 30$, the bent region
extends from $n=25$ to $n=45$ and $T = 0$. The speed was computed as 
$V=d/\Delta t$ where $\Delta t=50$ and $d$ is the distance travelled by the
polaron during this time interval.}\label{Fig2}
\end{figure}

Next we have investigated the dependence of the polaron dynamics on
$d\varphi$ - the gradient of the bending angle of the
chain. Fig. \ref{Fig2} shows the variation of the average velocity
$\langle V \rangle$ as a function of $d\varphi$. For our choice of 
parameter values the polaron does not move until a critical value
$d\varphi \approx 0.0056$ is  reached. The dependence of the velocity
on $d\varphi$ in the critical region follows a power law  $\langle V
\rangle \sim \left(d\varphi - d\varphi_{c} \right)^{\nu}$ akin to
elastic depinning of interfaces. In order  to express the average
speed in physical units, we have to  multiply the dimensionless speed
by $a/t \approx 328.5 \frac{nm}{ns}$.  Thus the average speed of the
polaron $\langle V \rangle \approx 16 nm/ns$.

In order to study the effect of thermal fluctuations on the  polaron
transport we have performed finite temperature $T\ne0$
simulations. For such simulations we generated a polaron on a straight
lattice at $T=0$ as above and saved it. Next, we deformed the lattice into 
a bent configuration as above and integrated our equations for $300$ 
units of time to ensure thermal equilibrium. We then restored the
electron field, without altering the phonon field, and numerically
integrated Eq (\ref{Polaron-Chain-Dynamics-Eq}) with the noise term to 
investigate the motion of the polaron along the chain. This simulation 
strategy, faithfully mimics the sudden excitation of a polaron and its time 
evolution.

\begin{figure}[tbp]
\centerline{\includegraphics[scale=0.4]{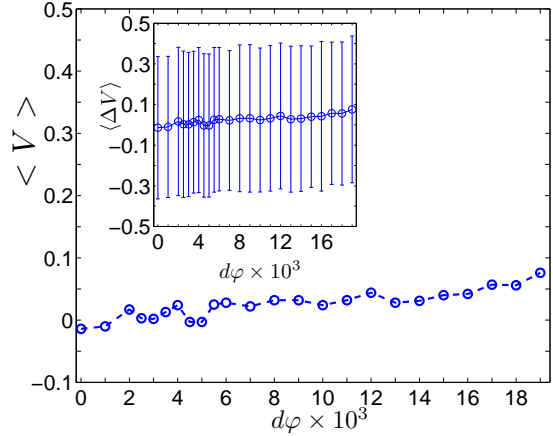}} \vspace*{-0.3cm}
\caption{Figure shows the variation of the average velocity $\langle V \rangle$ 
as a function of the bend angle $d\varphi$ for a $N=60$ chain. The chain is 
bent between $n=[25,45]$ nodes, and the initial position of the polaron is 
at $x_0=30$. The temperature $T=300 K$, and the data is averaged over 
$N_{run}=1000$ simulations. The speed was evaluated exactly like in figure 2.
Inset shows the variance of the velocity vs.\ 
$d\varphi$ data.}\label{Fig3}
\end{figure}

\begin{figure}[tbp]
\centerline{\includegraphics[scale=0.4]{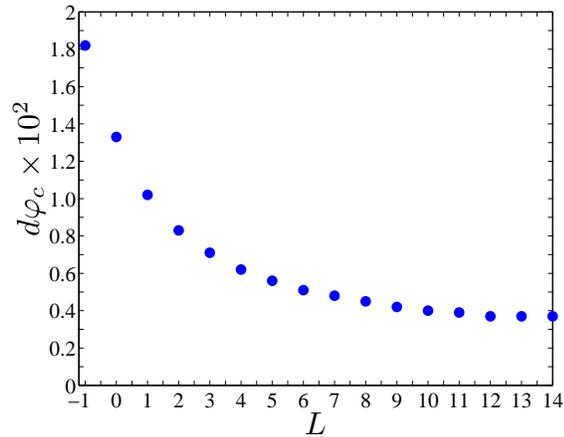}} \vspace*{-0.3cm}
\caption{Figure shows the variation of the critical bend angle $d\varphi_{c}$ 
as a function of the initial relative polaron position L with respect to the 
bent region, $n_{0,pol} = 25+L$. N = 60 and T = 0.
The speed was evaluated exactly 
like in figure 2.}\label{Fig4}
\end{figure}

Our finite temperature studies have focused on the polaron dynamics at
``room temperature'', \textit{i.e.} $T=300K$. Clearly, the  effects of
taking non-zero temperature are very significant as seen in
Fig. \ref{Fig3}. It is clear from this figure  that the polaron
undergoes large amplitude fluctuations about its initial position and
its velocity is significantly altered by the thermal  effects. The
thermally averaged polaron velocity shown in Fig. \ref{Fig3} is
averaged over $N_{run}=1000$ simulations, each time having measured
the average displacement over $t=50$ units of time. In all cases the life time 
of the polaron was of the order $t=200$, in our units, which corresponds to about 
$300ps$. Afterwards the electron was delocalised on the lattice. 
In Fig. \ref{Fig4} we have plotted the variation of the thermally
averaged velocity as a function of $d\varphi$; this is to be compared with the
curve obtained for $T=0$ in Fig. \ref{Fig2}.

Our simulations clearly show that the thermal effects lead to a wider 
displacement of the polaron wich can involve motion in both directions. The polaron is
thus subjected to a random walk motion in a system with an 
attractive force provided by the bend of the chain.
So, while the displacement of a polaron in a protein at room temperature is 
random, its position is still biased by the bending gradient; a 
thermalised polaron thus follows a random walk, but its average displacement
is similar to that of a polaron at $T=0$.

In order to quantify the dependence of the bend on the spontaneous
transport of polarons we  have calculated the dependence of the
critical incremental angle between segments $d\varphi_{c}$ on its
initial distance from the bend at $T=0$. This is shown  in
Fig. \ref{Fig4}. It is clear from the plot that $d\varphi_{c} \propto
\exp[- \xi L]$, where $\xi$ is a constant and L is the distance between the 
polaron and the edge of the region with the bend. 
\begin{figure}[tbp]
\centerline{\includegraphics[scale=0.4]{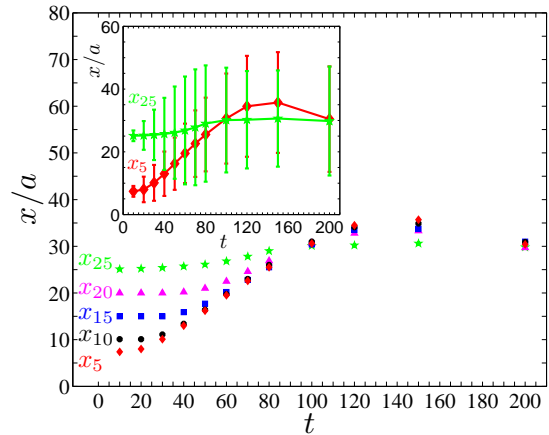}} \vspace*{-0.3cm}
\caption{Figure shows the average position of a polaron as a function of time 
for a straight chain of $N=60$ segments at $T=300K$ for different initial positions 
($x_i=5$ (red diamonds), $x_i=10$ (black circles), $x_i=15$ (blue squares), $x_i=20$ 
(magenta upper triangles) and $x_i=25$ (green stars) ) averaged over $N_{run}=1000$. 
Inset shows the average position and the fluctuation about the mean position of the 
polaron for two initial positions $x_i=5$ and $x_i=25$.}\label{Fig5}
\end{figure}

Finally  we have also explored a possible mechanism of the
energy/polaron transport for biomolecules \textit{i.e.} $\alpha$
helices in our case. Thus we have considered the transport of polarons
on straight chains at finite temperatures. Since the edges of the
chains  act as reflecting walls a polaron that is initially formed
close to one edge of the chain will get reflected  from it during the
course of its random walk motion with an amplitude set by the thermal
energy scale. Fig. \ref{Fig5}  shows the dependence of the mean
position of such a polaron that has initially been created at
$x_{0}=5, 10, \ldots 25$. While  the steady state polaron position is
the  same for all values of $x_{0}$, their early time behaviour is
significantly  different. Thus a polaron formed at $x_0 = 5$ has a high average 
velocity due to its reflections from the $n = 0$ edge and is, on average, 
transported further at a given time (say $t = 140$ ) than a polaron 
formed at $x_0 = 25$. This compels us  to conjecture the following
model of polaronic transport for biophysical systems: if a polaron is
generated by  electron-phonon interactions near an edge of a short
straight segment of a protein it will undergo reflections from the
nearby edge, leading to the directed motion towards the other edge. 
When the system contains receptor molecules that can absorb this non-linear
excitation and transfer it to other molecules or polymers, the polaron
displacement can be used to transport the energy released by ATP hydrolysis
to a nearby regions of the cell where it is needed.

\section{Conclusions}
\label{Conclude} 
In this paper we have studied the transport of energy along bio-polymers 
via polaronic mode. Starting from a model that describes coupled
electron phonon dynamics, we chose physically relevant parameter
values and studied polaron dynamics on bent configurations of
bio-polymers. We have found that the bent chain induces spontaneous
polaron transport due to some of its nodes being closer together and so 
generating an attractive potential for the polaron. When we looked at
this problem for proteins at zero temperature the movement of polarons was
observed when the gradient of the bend was quite small ($d\varphi < 0.0055$ in
our units). The polaron moved at a speed $V\approx 16nm/ns$ and lived for 
about $t=300ps$.  
The thermalisation of the polaron, at room temperature, modified some aspects
of the polaron transport: the polaron now exhibited a random walk motion
biased in the direction of the bending gradient. The bending gradient of the 
chain thus induced an average polaron displacement. Moreover, on a straight 
chain, we observed that a polaron near the edge of the chain exhibits 
spontaneous displacement in the direction away from the edge.
We have thus shown that polaron displacement can be induced spontaneously 
without the need to kick the polaron in one direction and that the direction
of transport can be determined by the configuration of the polymer.

\section{Acknowledgement}
\acknowledgments
BC was partially supported by EPSRC grant EP/I013377/1.
BP and WJZ were supported by the STFC research grant ST/H003649/1.


\begin{thebibliography}{0}

\bibitem{Davydov}
  \Name{Davydov A. S.}
  \REVIEW{J. Theor. Biol.}{38}{1973}{559};\Name{Davydov A. S.}\REVIEW{Phys. Scr.}{20}{1979}{387};\Name{Davydov A. S.}\REVIEW{Phys. D}{3}{1981}{1}.

\bibitem{Davydov:85} \Name{Davydov A. S.}
\Book{Solitons in Molecular Systems}
\Editor{D. Reidel}
\Publ{Dordrecht}
\Year{1985}; \Name{Davydov A. S.}\REVIEW{J. Theor. Biol.}{38}{1973}{559}.

\bibitem{Scott:92}
\Name{Scott A. C.}
\REVIEW{Phys. Rep.}{217}{1992}{1}.

\bibitem{Cruzeiro:09}
\Name{Cruzeiro L.}
\REVIEW{J. Biol. Phys.}{35}{2009}{43}.

\bibitem{Olsen:88}
\Name{Olsen O. H. \textit{et al.}}\REVIEW{Phys. Rev. A}{38}{1988}{5856};
\Name{Olsen O. H., Lomdahl P. S. \and Kerr W. C.}\REVIEW{Phys. Lett. A}{136}{1969}{402}.
 
\bibitem{LaMagna:95}
\Name{La Magna A., Pucci R., Piccitto G., \and Siringo F.}\REVIEW{Phys. Rev. B}{52}{1995}{273}

\bibitem{Zolotaryuk:96}\Name{Zolotaryuk V., Christiansen P. L., \and Savin A. V}
\REVIEW{Phys. Rev. E}{54}{1996}{3881}.

\bibitem{Christiansen:97}
\Name{Christiansen P. L., Zolotaryuk V., \and Savin A. V.}
\REVIEW{Phys. Rev. E}{56}{1997}{877}.

\bibitem{Henning:02}
\Name{Henning D.}\REVIEW{Phys. Rev. B}{65}{2002}{147302}. 

\bibitem{Kalosakas:98}
\Name{Kalosakas G., Aubry S., \and Tsironis G. P.}
\REVIEW{Phys. Rev. B}{58}{1998}{3094}.

\bibitem{Voulgarakis:01}
\Name{Voulgarakis N. K \and Tsironis G. P.}
\REVIEW{Phys. Rev. B}{63}{2001}{014302}.

\bibitem{Eremko04}
\Name{Brizhik L., Eremko A., Piette B. \and Zakrzewski W. J.}
\REVIEW{Phys. Rev. E}{70}{2004}{031914}.

\bibitem{Brizhik10}
\Name{Brizhik L., Eremko A., Piette B. \and Zakrzewski W. J.}
\REVIEW{J. Phys. : Condens. Matter.}{22}{2010}{155105}

\bibitem{Ming02}
\Name{Mingaleev S. F., Gaididei Y. B., Christiansen P. L. \and Kivshar Y. S.}
\REVIEW{Europhys. Lett.}{59}{2002}{403}.

\bibitem{Scott82}
\Name{Scott A. C.}
\REVIEW{Phys. Rev. A}{26}{1988}{578}.

\bibitem{Phillips1996}
\Name{Phillips G. N., Chacko S.}
\REVIEW{Biopolymers}{38}{1996}{89}







\end{thebibliography}
\end{document}